\documentclass[12pt]{iopart}
\usepackage{amssymb}
\usepackage{palatino}
\usepackage{graphicx}
\newcommand {\tobs}{t_{\mathrm{obs}}}
\newcommand {\Tc}{T_{\mathrm{c}}}
\newcommand {\Sc}{s_{\mathrm{c}}}

\begin{document}

\title[Thermodynamics of trajectories of 1$d$ Ising model]{Thermodynamics of trajectories of 
the one-dimensional Ising model}

\author{Ernesto S Loscar\footnote {Permanent address: Instituto de Investigaciones 
                                    Fisicoqu\'{\i}micas Te\'{o}ricas y Aplicadas 
                                    (INIFTA), UNLP, CCT-La Pata, CONICET,
                                    Suc.4, CC 16, (1900) La Plata, Argentina.}
, Antonia S J S Mey and Juan P Garrahan}

\address{School of Physics \& Astronomy, University of Nottingham, NG7 2RD
UK}
\ead{yasser.loscar@gmail.com, ppxasjsm@nottingham.ac.uk,\\juan.garrahan@nottingham.ac.uk }
\begin{abstract}
We present a numerical study of the dynamics of the one-dimensional Ising model
by applying the large-deviation method to describe ensembles of dynamical
trajectories.  In this approach trajectories are classified according to a
dynamical order parameter and the structure of ensembles of trajectories can be
understood from the properties of large-deviation functions, which play the role of
dynamical free-energies. We consider both Glauber and Kawasaki dynamics, and
also the presence of a magnetic field.  For Glauber dynamics in the absence of a
field we confirm the analytic predictions of Jack and Sollich about the
existence of critical dynamical, or space-time, phase transitions at critical
values of the ``counting'' field $s$.  In the presence of a magnetic field the
dynamical phase diagram also displays first order transition surfaces. We
discuss how these non-equilibrium transitions in the 1$d$ Ising model relate to
the equilibrium ones of the 2$d$ Ising model.  
For Kawasaki dynamic we find a much simple dynamical phase structure, with
transitions reminiscent of those seen in kinetically constrained models.

\end{abstract}

\maketitle

\section{Introduction}

In this paper we perform a detailed numerical study of the trajectories of the one-dimensional 
Ising model under both Glauber and Kawasaki dynamics.  We do so by means of the 
``thermodynamics of trajectories'' method introduced in references
\cite{Merolle2005,Lecomte2007,Garrahan2007,Garrahan2009}, which in turn is based on
Ruelle's thermodynamic formalism for dynamical systems~\cite{Ruelle1978,Gaspard1998}. 
The basic idea is to understand dynamical properties of a many-body system by treating ensembles 
of trajectories in a way which is analogous to how one understands static or structural 
properties via the ensemble method for configurations (or microstates) in standard equilibrium 
statistical mechanics. In order to classify trajectories one defines a suitable dynamical order
parameter, one that captures the essence of the physics one is trying to uncover. As we will 
show below, a convenient one is given by the ``dynamical activity''
~\cite{Lecomte2007,Garrahan2007,Maes2008}, which for a lattice problem like the
ones we study in this paper corresponds to the number of configuration changes
(spin-flips in our case) in a trajectory. Alternatively, it could be the time integral of a 
standard observable, such us the time-integral magnetization. One can then study ensembles of
trajectories classified by dynamical activity, or more conveniently as we show
below, by its conjugate ``counting'' field $s$.  For trajectories observed over
long-times a large-deviation principle~\cite{Touchette2009} holds such that the
distributions of the activity over the ensemble of trajectories, or their
corresponding generating functions, are determined by time independent
``large-deviation'' functions, which play a role analogous to thermodynamic
potentials (entropies or free-energies) for the dynamics.  This approach has
proved successful recently in uncovering dynamical or ``space-time'' phase
transitions in a variety of classical many-body systems, most notably glass
models, see e.g.~\cite{Garrahan2007,Hedges2009,Lecomte2010,Elmatad2010}.
In these systems thermodynamics is either trivial, as in kinetically constrained models
of glasses~\cite{Garriga2005}, or not very revealing, as in more realistic liquid
models~\cite{Kob-Binder}, but the dynamical phase structure is extremely rich: i.e.
complex and highly correlated relaxational dynamics can emerge independently of
any form of complex thermodynamics, and this is clearly revealed by the
large-deviation method for trajectories. 

So a natural question is the following: what is the dynamical phase structure of systems 
which do display strong thermodynamic features, and is there any relationship between 
thermodynamic and dynamical, or space-time, phases? Recent work has addressed this issue 
in certain simple models, including mean-field Ising and Potts models~\cite{Bodineau2008}
and certain mean-field spin glass models~\cite{Duijvendijk2010,Jack2010a}. Specifically, 
and directly related to the work we present here, there is a recent paper by Jack and 
Sollich~\cite{Jack2010} in which they study analytically the dynamical phases of the 
1$d$ Ising model with Glauber dynamics. By mapping the problem to that of a 
quantum Ising model in a transverse field, they managed to obtain the large-deviation 
function that plays the role of a free-energy for trajectories, and showed that in an enlarged $T-s$
(temperature/counting-field) space of parameters there are second-order
dynamical transitions between active (and paramagnetic) and low-activity (and
ferromagnetic) dynamical phases---transitions which are not observed under
standard thermodynamic conditions---with critical properties of the classical 2$d$
Ising universality class.  

The current paper builds on Jack-Sollich's results.  By means of path sampling
methods~\cite{Dellago2002} we generate ensembles of trajectories biases by the counting field
$s$, the so-called $s$-ensembles~\cite{Hedges2009}.  We confirm Jack-Sollich's predictions 
for Glauber dynamics in the absence of the magnetic field, providing also detailed analysis of the 
scaling properties of the dynamical transitions. We also extend the problem to the case 
of non-zero magnetic field.  This allows us to map out in detail the full dynamical phase 
diagram, revealing a host of first- and second-order dynamical transitions, and the connection 
between dynamical and thermodynamical phases.  We also consider the case of Kawasaki
dynamics and show that the dynamical phase structure in this case is analogous
to that of kinetically constrained models of glasses~\cite{Garrahan2007,Garrahan2009}.

This paper is structured as follows: first, in \sref{section:formalism},
we will introduce the theoretical framework and large deviation formalism, followed 
by two main sections. In \sref{section:Glauber} the theoretical and computational 
results of the 1$d$ Ising model using Glauber dynamics are discussed with strong focus on 
the extension of the phase diagram through an external magnetic field and the resulting 
scaling behaviour. In the \sref{section:Kw} we will look at the dynamic behaviour 
when applying Kawasaki dynamics. Finally our conclusions are given in~\sref{section:Conclusions}.

\section{Theory and Methodology}{\label{section:formalism}}

The objective of this paper is a computational study of the dynamical processes of a $1d$ Ising 
spin chain in the presence and absence of a magnetic field, and within Glauber or Kawasaki 
dynamics. In order to achieve this, we apply what is referred to as the $s$-ensemble in the 
literature~\cite{Garrahan2007, Garrahan2009}, which will be briefly reviewed in the following. 
In order to study the dynamics, or time evolution of the system of interest, an appropriate 
observable capturing the dynamics needs to be defined, as well as a set length of time for which 
the system is observed. In this way a trajectory or history of the system can be defined with 
a given length in time $\tobs$.  The trajectory is the sequence of
configurations, and thus contains information about the instantaneous state of
the system at each point in time. Any configurational change in the system gives
rise to a dynamic observable.  The most basic one is the number of spin flips in
a given trajectory. This is referred to as the activity $K$~\cite{Lecomte2007,Garrahan2007,Maes2008}.
In the thermodynamic Gibbs ensemble the probability of observing a specific configuration is governed by 
the partition function $Z$. For the dynamics an equivalent object can be constructed, based on the 
formalism by Ruelle~\cite{Ruelle1978}, which gives information about the probability of observing 
a given trajectory.  The dynamic partition function is defined as 
\begin{equation}
Z(s,\tobs) = \langle e^{-sK} \rangle_0~~~,
\end{equation}
where $K$ is the activity of the system. We are considering path averages, denoted by $\langle~\rangle_0$,
over a set time intervals starting at $t=0$ until a later time $t=\tobs$ in an equilibrated system.
The parameter $s$ is a biasing or counting field conjugate to $K$ and hence allows subtle control over the 
ensemble of trajectories.  In the context of full counting statistics in electronic transport 
such parameters are referred as ``counting fields'' ~\cite{Levitov1996}. 
It plays an equivalent role to that of the inverse temperature $\beta$ as
conjugate field of the energy (or to the magnetic field as a conjugate to the
magnetization in a magnetic problem) in the Gibbs ensemble. 

In order to give a physical meaning to this space-time analogue of the partition 
function, the large deviation formalism is used. Traditionally the thermodynamic limit, as in 
an infinite system size, is considered. In this case the limit of very large observational
time $\tobs$ will be considered. Thus the partition function can be written in form of:
\begin{equation}\label{eq:Z}
Z(s,\tobs) \sim e^{\tobs\psi(s)}~~,
\end{equation}
where the ``large-deviation function'' $\psi$~\cite{Lecomte2007,Garrahan2007,Touchette2009}
plays the role of a free-energy function for trajectories. The structure of this function 
provides information about the  dynamic phase space behaviour of the system,
e.g. it singularities indicate phase transitions in ensembles of trajectories. 

In order to analyse the dynamic behaviour we consider a set of time extensive observables, well 
defined within trajectories, such as the time integrated magnetization or energy for 
the case of an Ising magnet. If $A$ is one of these observables, its expectation
values biased according to the modified path ensemble is given by
\begin{equation}
\langle A \rangle_s=\frac{\langle A e^{-sK}\rangle_0}{\langle
e^{-sK}\rangle_0}~~.
\end{equation}
Note that for $s=0$ the equilibrium expectation is recovered. Hence a set of time 
intensive variables can be defined, normalized by the choice of observational time 
and system size $N$. For example for the intensive activity $k$ we have
\begin{equation}
k(s) = \frac{\langle K \rangle_s}{N\tobs}~~.
\end{equation}
This leads to a generalization of the susceptibilities associated to this
ensemble, which are defined in the usual way

\begin{equation}\label{Eq:activitysusc}
\chi_{_\texttt{k}} = \frac{\langle K^2 \rangle_s - \langle K \rangle_s
^2}{N\tobs}~~.
\end{equation}
This will further allow the study of the critical behaviour within the dynamics.

This theoretical framework will be used in order to study a $1d$ Ising spin chain, on a lattice 
of size $N$, under the presence of an external magnetic field. The general Hamiltonian for
such a system is given by:
\begin{equation}\label{equation:Ising-1d}
  {\cal H}=-J\displaystyle\sum_{i=1}^{N} \sigma_i\sigma_{i+1}
  -h\displaystyle\sum_{i=1}^{N}\sigma_i ,
\end{equation}
where the spins $\sigma_i$ take values of $1$ or $-1$, depending on their direction, and $J>0$ is the 
coupling constant between first neighbours, and takes the value of 1 for our purposes.
The second sum describes the coupling of individual spins to the external magnetic field $h$.
Simulations are done using a Monte Carlo scheme in which periodic boundary conditions are considered 
for the spatial dimension. As usual, the Monte Carlo time is taken as one unity when $N$ attempts of 
spin flips are performed. The trajectory length $\tobs$ is predefined. Adjusting the field $s$,
will allow biasing towards lower or larger activity and thus samples the tails of the equilibrium 
distribution. Rather than just running different realizations of the equilibrium distribution, we use 
an adaptation of transition path sampling (TPS) with $s$-ensemble biasing~\cite{Dellago2002}, 
similar to the one used to generate $s$-ensembles in certain glassy models~\cite{Hedges2009,Elmatad2010}. 
Once an equilibrium trajectory is produced, this can be used as blue print for the next trajectory. 
As in TPS only part of the original trajectory is used. A shooting point is chosen, from which the 
system is propagated in time, until the chosen observational time is reached. Then, for the original and 
new trajectory the activity is calculated, which is the incremental number of spin flips observed within
the length of the trajectory. This leads to the acceptance of the newly proposed trajectory according 
to the standard Metropolis criterion:
\begin{equation}\label{eq:acceptance}
P_{\mathrm{accept}} = \min(1, e^{-s\Delta K})
\end{equation}
where $\Delta K$ is the difference in activity between the two trajectories. The employed TPS method 
allows the generation of trajectories with similar values of $K$ and therefore ensures that new 
trajectories are likely to be accepted. This method has been employed for the generation of 
data in this paper in an efficient way.

In terms of the Monte Carlo dynamics used within this work, there were two separate computational 
studies, one using Glauber dynamics and the other Kawasaki dynamics. The motivation for
the use of Glauber dynamics over the generally preferred Metropolis dynamics is theoretically 
motivated. A recent paper has a theoretical solution for the space time behaviour of the $1d$ Ising 
model with Glauber dynamics, employing an ensemble similar to the $s$-ensemble presented 
above~\cite{Jack2010}. Jack and Sollich refer to this ensemble the $g$-ensemble. 
Rather than biasing towards activity an (integrated) energy bias is introduced,
where $g$ is the conjugate field to the energy. Analogously to the $s$-ensemble a dynamical 
free energy $\phi(g)$ can be defined for this ensemble. Only in terms of the Glauber dynamics a 
theoretical solution is readily available in order  to quantify the critical behaviour of the system. 
Thus, the computational model uses a Glauber acceptance criterion, but for the Kawasaki dynamics 
the preferred Metropolis criterion will be used. This will be presented in detail in the following 
two sections.


\section{Glauber Dynamics}\label{section:Glauber}

In this section we will present the space-time phase diagram obtained from our
computational results using Glauber dynamics.  In terms of the evolution of the
Monte Carlo dynamics, the probability of accepting a spin flip is given
according to the Glauber acceptance criteria~\cite{Glauber1963}
\begin{equation}
P_{\mathrm{accept}}=\frac{1}{1+e^{\beta\Delta E}}
\label{eq:Glauber}
\end{equation}
where $\beta$ is the inverse temperature $\beta = 1/k_BT$ and the Boltzmann
constant $k_B$ 
is assumed to be 1. 

\subsection{The phase diagram}\label{section:phase-diagram}

Through the temporal evolution a dynamic phase diagram can be constructed in temperature and biasing field $s$. 
Reference~\cite{Jack2010} considers a modified path ensemble equivalent to the discussed $s$-ensemble. 
In the $g$-ensemble, similar to the dynamic free energy $\psi(s)$ introduced in~\sref{section:formalism}, 
the dynamical free energy ${\phi}(g)$ of the biased ensemble of trajectories can be obtained using 
the large deviations of the thermodynamic energy with zero magnetic field. It was further shown, that the 
second derivative of $\phi(g)$ diverges to a line of critical points, which can be shown to belong to the 
universality class of the 2$d$ Ising model. An equivalence between the activity constrained ensemble 
(with biased field $s$) and the energy constrained ensemble (with biased field $g$) can be established, 
since for the Ising problem we are considering biasing the energy is akin to biasing in terms of the escape
rate of the corresponding continuous-time Markov chain, see e.g.~\cite{Garrahan2009}. In this model, due 
this equivalence the free energy of the $s$-ensemble ${\psi}(s)$ shares the same critical properties of 
${\phi}(g)$, because both functions only differ in a regular factor. Therefore the dependence on $s$ and 
$T$ of the second order critical line is given by the following relationship~\cite{Jack2010}
\begin{equation}\label{eq:critical-line}
s(T)=-\ln \left[\tanh \left(2 J /k_BT\right)\right].
\end{equation}

\begin{figure}

\includegraphics[scale = 0.7]{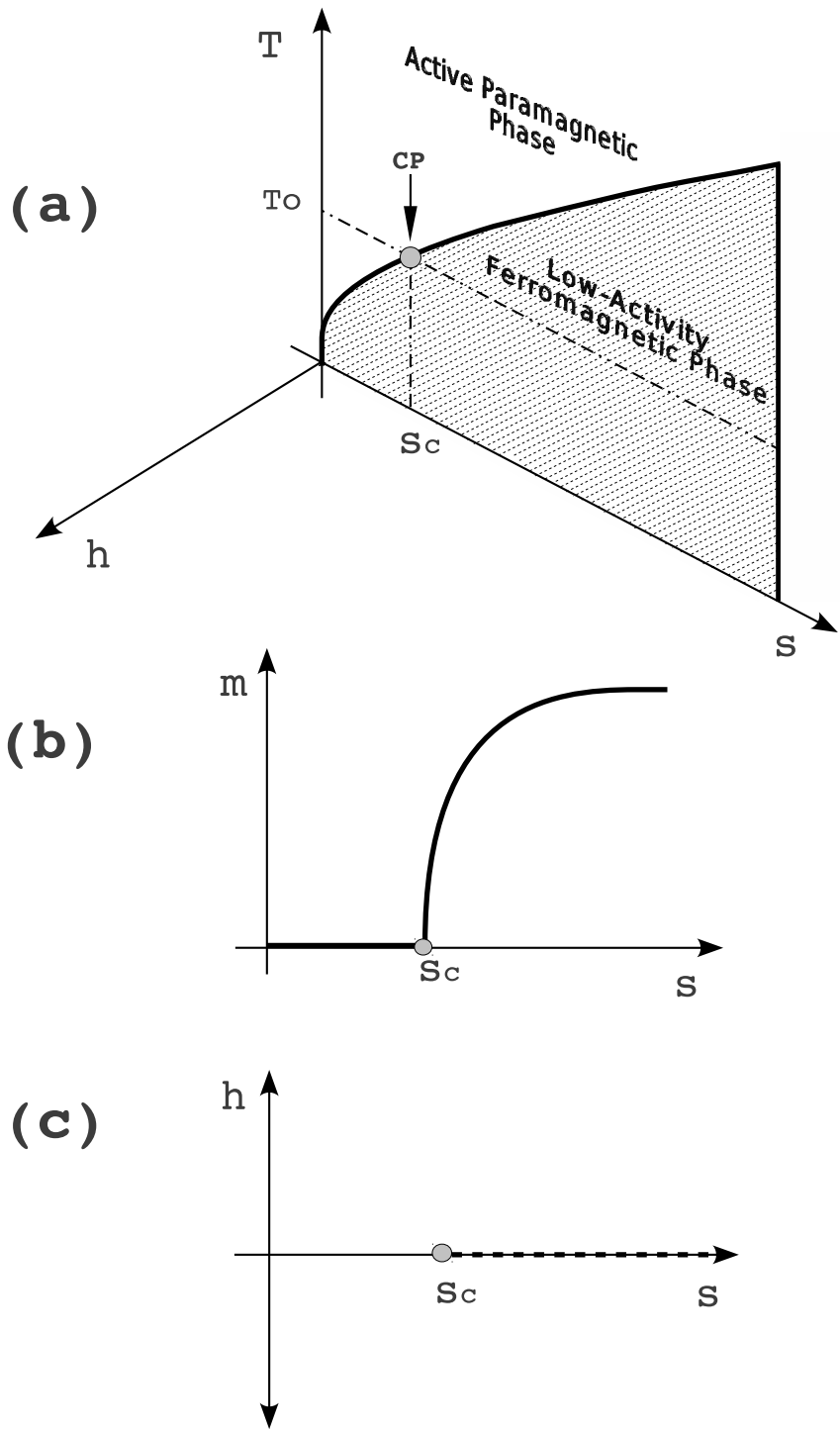}
\caption 
        {Schematic representation of the phase diagram of the Glauber 1$d$ Ising model in the $s$-ensemble with 
         magnetic field $h$. (a) The phase space is defined through three variables $T,s,$ and $h$. Here the 
         continuous line, into the plane $h=0$, represents the critical points given by the theoretical solution 
         in reference~\cite{Jack2010} separating paramagnetic (and active) and ferromagnetic 
         (and low-activity) dynamical phases. 
         The dashed line correspond to the conditions $T=T_0=\mathrm{const}$ and 
         $h=0$ and contains a critical point (CP) marked by a circle. (b) Spontaneous magnetization as a function of 
         $s$ at temperature $T=T_0$ and $h=0$, (dashed line drawn in  (a)). The symmetry is broken by the biasing 
         field $s$ for $s>\Sc$ just at the critical point $s=\Sc$. (c) The plane $h,s$ taking $T=T_0$. The 
         critical point $s=\Sc$ divides the line in two phases, a paramagnetic phase for $s<\Sc$  and a ferromagnetic
         phase $s>\Sc$. For $s>\Sc$, the dashed line represents expected points of first order transitions. 
        }
\label{Fig:phase-diagram}
\end{figure}

\Fref{Fig:phase-diagram} shows, a schematic representation of the dynamic phase diagram for 
the Ising chain with Glauber dynamics. This phase diagram is defined in a $3d$ parameter space 
depending on the set of variables $\{s,T,h\}$. In particular the line in~\fref{Fig:phase-diagram}(a) 
represents the curve given by \Eref{eq:critical-line}, which is a line of critical points 
in the plane of $h=0$, dividing this plane into two phases: the paramagnetic phase (for $s<\Sc$)
and the ferromagnetic phase (for $s>\Sc$)~\cite{Jack2010}.

\Fref{Fig:phase-diagram}(a) also shows a line for zero magnetic field given by isothermal 
conditions $T=T_0$ (dashed line). As indicated, this line contains a critical point corresponding 
to $\Sc=s(T_0)$ given by~\Eref{eq:critical-line}.~\Fref{Fig:phase-diagram}(b) shows
the magnetization of the Ising chain as a function of the biasing field $s$. Here the critical point 
$s=\Sc$ separates the ferromagnetic phase for $s>\Sc$ with non zero spontaneous 
magnetization $m\neq0$ and, the paramagnetic phase for $s<\Sc$ with $m=0$. Tuning the biasing field $s$, once the critical point is reached, 
the symmetry of the system is spontaneously broken.
The inclusion of the magnetic field in the isothermal behaviour gives rise to a third dimension and 
therefore the plane $T=T_0$ can be considered as shown in the~\fref{Fig:phase-diagram}(c).  
Here for $s<\Sc$ the magnetic field acts on a paramagnet, while for $s>\Sc$ the magnetic field acts on a 
ferromagnet.~\Fref{Fig:phase-diagram}(b) and~\fref{Fig:phase-diagram}(c) are analogous to those 
of the 2$d$ Ising model by assuming $s \rightarrow \beta$. In fact by taking $s=\Sc$ and using the 
temperature $T$ (instead of $s$) as the control parameter, the critical behaviour is in complete analogy 
with the 2$d$ Ising model. In the presence of an external magnetic field $h$ we expect the system to exhibit 
a first order phase transition along the dashed line showed in the~\fref{Fig:phase-diagram}(c) 
where $s>\Sc$. This obviously gives rise to an entire surface of critical points for $h=0$ 
below the curve of $s(T)$ as showed in~\fref{Fig:phase-diagram}(a). In the following, the second 
order critical behaviour as predicted by the theory of~\Eref{eq:critical-line} will be studied through Monte 
Carlo simulations as well as the validity of the existence of the first order surface by means of the presence 
of an external magnetic field.

\subsection{Dynamical continuous phase transition}

In this section we study by means of Monte Carlo simulation the critical behaviour of the critical
points given by the~\Eref{eq:critical-line}. As was established in the discussion of 
the~\sref{section:phase-diagram} describing the~\fref{Fig:phase-diagram}, we can use 
a $s$-fixed curve or a isothermal in order to reach the critical point. 
Trajectories are generated as described in~\sref{section:formalism} using a fixed temperature
$T$ and with zero magnetic field. From each accepted trajectory, of a given value of the biasing parameter 
$s$, the set of time extensive variables is obtained. These are the activity $K$ as the number of the 
changes of configuration, the integrated energy $E=\int_{0}^{\tobs} dt' u(t')$, and the integrated magnetization 
$M=\int_{0}^{\tobs} dt' m(t')$, where $m(t)=\sum_{i=1}^{N}s_i(t)$. 

In order to study critical properties, we need to apply the finite size scaling theory.
This is a commonly used technique in computational physics~\cite{Landau2005}. 
As we are investigating the scaling behaviour of dynamic properties, 
the temporal size is given by the observational time $\tobs$. In fact, we can see from 
the definition given by the \Eref{eq:Z} that it is analogous to the spatial system size 
when the equilibrium thermodynamic free energy is defined.
Of course, in our simulations we have to use a system with finite spatial size, i.e. the 
number of spins $N$. 
We have observed that the behaviour of the system is very sensitive with respect to the length of the 
trajectories $\tobs$ (much more than to the length $N$ of the chain). 
In order to have a consistent analysis of the finite size effects, we have chosen to study the behaviour 
considering $\tobs$ as the scaling parameter, so that the lattice size $N$ is set to fixed 
value.

It is worth mentioning that the dynamic evolution of the this system can be described with a time 
evolution operator, which is equivalent to a quantum Hamiltonian. From this point of view the length 
($\tobs$)~of the trajectories is equivalent to the inverse of the temperature $T^\texttt{(q)}$,
as in the associated quantum problem discussed in~\cite{Jack2010} 
(so that $\tobs~\leftrightarrow  1/T^\texttt{(q)}$).
It is well known that there is a mapping between the classical 2$d$ Ising model and the 
quantum Ising chain~\cite{Sachdev1999}. From this mapping we know that the inverse of $T^\texttt{(q)}$ 
is equivalent to the size in one spatial direction in the 2$d$ Ising model 
(that is $ 1/T^\texttt{(q)} \leftrightarrow L$). Therefore we conclude that $\tobs$ in 
our problem is equivalent to a linear size $L$ in the classical 2$d$ Ising model (i.e. $\tobs~ \leftrightarrow L$).
According to this we assume that the finite size effects are given by the same well known relationships 
used for the thermodynamic 2$d$ Ising model, where we now replace $L$ with $\tobs$. 

To determinate the critical temperature of a continuous phase transition we can use the peak 
of the susceptibilities. According to the previous discussion we expect that susceptibilities 
have the scaling forms given by \cite{Landau2005}
\begin{equation}\label{eq:FSSmax}
\chi(\tobs) = \tobs^{\alpha / \nu} f \left( (T-\Tc) \tobs^{1/\nu} \right)~~,
\end{equation}
where $\alpha$ is the exponent that characterizes the divergence of the susceptibility $\chi$ and
$\nu$ the length correlation exponent. Furthermore, the location of the peaks define an effective 
size-dependent transition temperature which is expected to vary as
\begin{equation}\label{eq:FSSextrapolation}
\Tc(\tobs) = \Tc(\infty) + A \tobs^{-1/\nu} ~~ .
\end{equation}

\begin{figure}
\includegraphics[width =.9\columnwidth]{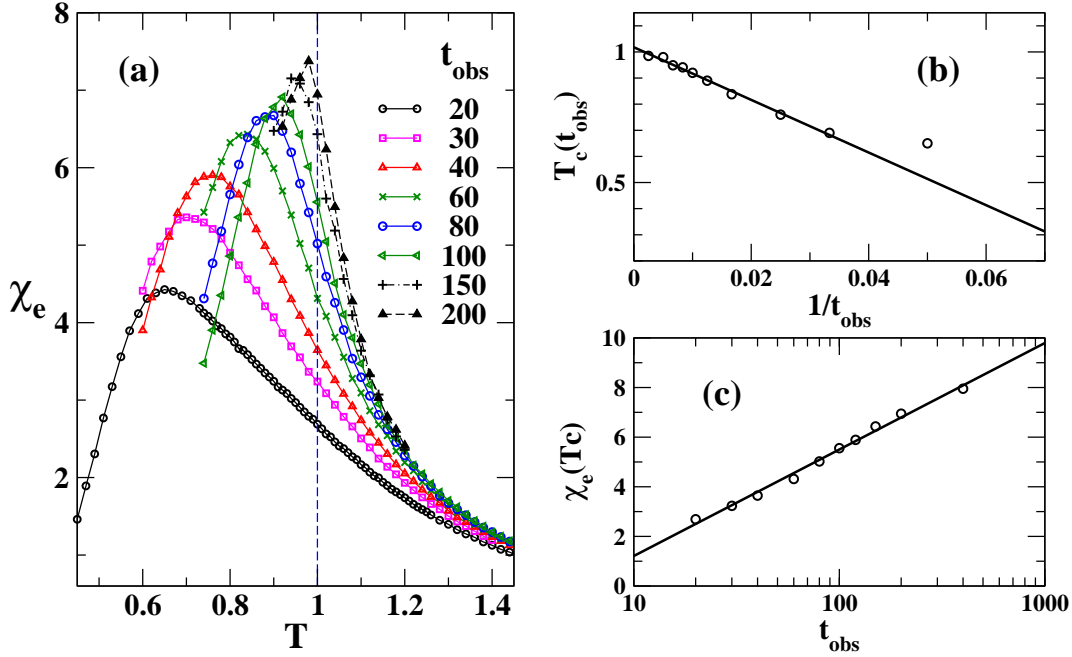}
\caption 
	{(a) Susceptibility of the energy as a function of $T$ at constant $s=0.036~635~37$.
             Vertical dashed line corresponds to $T_c=1.00$ given by \Eref{eq:critical-line}.  
         (b) Susceptibility at $T_c=1.00$ versus $\tobs$. The divergence is logarithmic, as in the
             case of the specific heat in the 2$d$ Ising model. 
         (c) Effective critical $s$, taken as the $s$ at the peak of~\fref{Fig:energy}
             (a), against $t_{obs}$. The extrapolated value, for $\tobs\rightarrow \infty$, 
             is $\Tc=1.02\pm0.02$. 
	}
\label{Fig:energy}
\end{figure}

As it is showed in the~\fref{Fig:phase-diagram}, we can use an $s$-fixed curve or an isothermal  
one in order to reach the critical point, and therefore the scaling given by the \Eref{eq:FSSextrapolation}
is expected in terms of $T$ or in terms of $s$, respectively. 

\Fref{Fig:energy} shows the simulation results for a fixed value of $s$ and varying the
temperature $T$. We have chosen the value of $s$ as $\Sc = 0.036~635~37$, which correspond 
to $\Sc$ at $s(T\equiv1)$ as given by~\Eref{eq:critical-line}. 
We define the generalized susceptibility of the energy $\chi_{_\texttt{e}}$ by means of

\begin{equation}\label{Eq:energysusc}
\chi_{_\texttt{e}} = \frac{\langle E^2 \rangle_s - \langle E \rangle_s ^2}{T^2N\tobs}~~.
\end{equation}
In~\fref{Fig:energy}(a) we plot $\chi_{_\texttt{e}}$ as a function of the temperatures and 
for different trajectory lengths $\tobs$. We used a lattice size $N=64$, and the averages 
were taken over $5\times10^6$ realizations of single trajectories. 
Peaks close the critical point are clearly observed, which sharpen, as the system size is increased 
(i.e. $\tobs$). We have used the location of these peaks as the definition of a 
size-effective ``critical temperature'' $\Tc(\tobs)$. 
  
~\Fref{Fig:energy}(b) shows the effective critical temperature as a function of the inverse of the 
trajectory length $\tobs$. We can see that the behaviour is compatible with~\Eref{eq:FSSextrapolation} 
taking an exponent $\nu=1$ corresponding to the 2$d$ Ising universality class. 
The continuous line is given by the linear fit of the data. From this our extrapolated estimated value 
for the critical temperature is $\Tc=1.02\pm0.02$, and it is consistent with the theoretical value given 
by~\Eref{eq:critical-line}. Also,~\fref{Fig:energy}(c) shows the value of the susceptibility 
measured in~\fref{Fig:energy}(a) for $T=\Tc=1$ and different sizes in log-linear scale.

As the dynamical free energy $\phi(g)$ must have the same critical properties as the thermodynamic free 
energy $f(T)$ of the 2$d$ Ising model, its second derivative must be analogous to the behaviour of the 
specific heat. In fact it can be shown that the second derivative of $\phi(g)$ has a logarithmic divergence 
when the critical point is approximated~\cite{Jack2010}. \Fref{Fig:energy}(c) shows that a logarithmic 
behaviour is observed, which means that~\Eref{eq:FSSmax} holds with an exponent $\alpha=0$, according
to the previous sentences. 

\begin{figure}
\includegraphics[width=1.0\columnwidth]{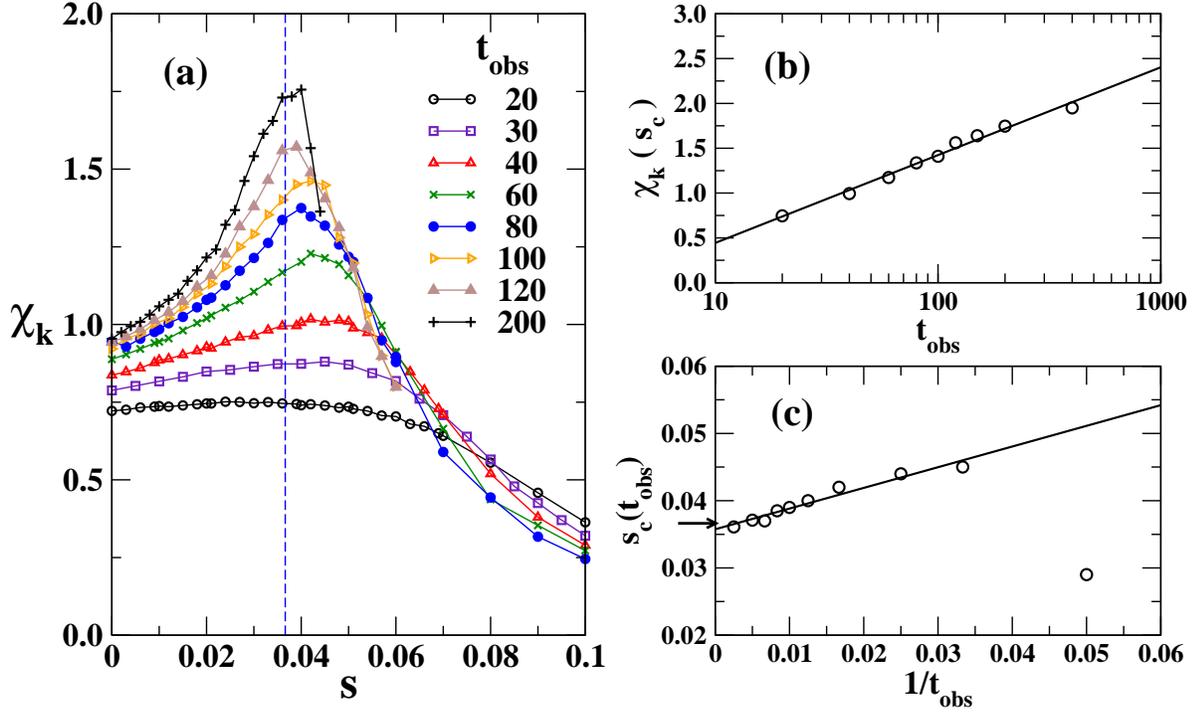}
\caption 
	{(a) Susceptibility of the activity as a function of $s$ at constant temperature $T=1$.
             Vertical dashed line corresponds to $s=\Sc=0.036~635~37$ given by \Eref{eq:critical-line}.  
         (b) Convergence of the effective critical $s$, taken as the s at the peak of figure left,
         in $\tobs$. The extrapolated value, for $\tobs\rightarrow \infty$, is $\Sc=0.036\pm0.001)$. 
         The arrow indicates the theoretical value from~\Eref{eq:critical-line}
         (c) Susceptibility at $s=\Sc=0.036~635~37$ constant versus $\tobs$. The critical divergence 
         of the $\chi_{\texttt{k}}$ is logarithmic. 
	}
\label{Fig:chi_k}
\end{figure}

As we have already indicated, the critical behaviour can also be accessed by choosing a constant temperature $T_0$ 
and considering different values of the biasing field $s$. In~\fref{Fig:chi_k}(a) we plot the susceptibility 
of the activity $\chi_{\texttt{k}}$, given by the~\Eref{Eq:activitysusc}, versus $s$ with fixed $T=1$.
The observed behaviour is analogous to the energy fluctuations. Again the value of the critical 
$\Sc$ can be extrapolated making use of~\Eref{eq:FSSextrapolation},
but in this case using the biasing field $s$ instead of $T$. In~\fref{Fig:chi_k}(b) we 
plot the effective critical $s$ and the linear fit gives us an estimate of the critical 
field, that is $\Sc=0.036\pm0.01$. Once again this result is consistent with $\nu=1$. 
Also~\fref{Fig:chi_k}(c) shows the susceptibility $\chi_{\texttt{k}}$ for $s=0.036~635~37$ in a log-lineal 
scale. We observe again that the~\Eref{eq:FSSmax} is valid with a logarithmic divergence of the 
susceptibility of the activity, that means $\alpha = 0$. 

So far a complete equivalence, in the critical behaviour, between the biasing field $s$ and the inverse 
of the temperature $1/T$ is observed. Also, similar results are obtained when studying finite size effects 
of the fluctuations of $K$ (or $E$) with respect to $T$ (or $s$).

Let us now consider the system in the presence of an external magnetic field $h$, which couples to 
the spins of the chain according to~\Eref{equation:Ising-1d}. We have studied the Ising Glauber chain 
with the magnetic field $h$ close to the critical point. 

We remark that the magnetic susceptibility of the 2$d$ Ising model is given by
\begin{equation}\label{eq:chiT_Ising}
\chi_{_\texttt{T}} =\frac{\langle m^2\rangle-\langle m\rangle ^2}{kTN}= \frac{1}{kTN}
\sum_{i,j=1}^{N} 
G(r_i,r_j)~~, 
\end{equation} 
where $G(r_i,r_j)= \langle \sigma_i \sigma_j \rangle - \langle \sigma_i \rangle  \langle \sigma_j \rangle$ 
is the two point correlation function, so that $\chi_{_\texttt{T}}$ is static, and is a measure 
of the spatial correlations. In the critical isotherm, the magnetic susceptibility 
shows a critical divergence given by
\begin{equation}\label{eq:Delta}
\chi_{_\texttt{T}} \propto  h^{-(1-1/\delta)}
\end{equation}
being the critical exponent $\delta=15$, thus $1-1/\delta=14/15\approx0.933$. 

Following the definition of the generalized dynamic susceptibilities, we define the time-spatial 
magnetic susceptibilities as
\begin{equation}\label{Eq:chi_correlation}
\chi_{_\texttt{m}} =\frac{\langle M^2\rangle-\langle M\rangle ^2}{kTN\tobs}= \frac{1}{kTN\tobs}
\sum_{i,j=1}^{N} \int_{0}^{\tobs} \int_{0}^{\tobs} dt''dt'
G(r_i,r_j,t',t'')~, 
\end{equation} 
where now 
$G(r_i,r_j,t',t'')= \langle \sigma_i(t')\sigma_j(t'') \rangle_s - \langle \sigma_i(t') \rangle_s  \langle  \sigma_j(t'') \rangle_s$
is the two point and two time correlation function. This susceptibility ($\chi_{_\texttt{m}}$)
has both spatial and temporal correlations. We can also define in the $s$-ensemble the static 
(or instantaneous) susceptibility in the same way as in~\Eref{eq:chiT_Ising}, which
is given by
\begin{equation}\label{eq:chiT_Ising-s}
\chi^{(s)}_{_\texttt{T}} =\frac{\langle m^2\rangle_s-\langle m\rangle_s ^2}{kTN}
= \frac{1}{kTN}
\sum_{i,j=1}^{N} G(r_i,r_j,\tau=0)~~, 
\end{equation} 
where we have introduced $\tau \equiv \arrowvert t''-t' \arrowvert$.

We can apply the schema used for standard critical phenomena given by the dynamic scaling hypothesis,
and obtain a relationship between the critical exponents corresponding to these susceptibilities 
\cite{GoldenfeldBook}.
So, in the thermodynamics limit, assuming a translational and temporal invariant correlation 
function $G$, and the existence of a diverging correlation length $\xi$, we obtain
\begin{equation}\label{eq:chisT}
\chi^{(s)}_{_\texttt{T}} \varpropto
\sum_{R} G(R,0) \varpropto \xi^{2-\eta} ~~, 
\end{equation} 
and 
\begin{equation}\label{eq:chiM}
\chi_{_\texttt{m}} \varpropto 
\sum_{R} \int d\tau'
G(R,\tau') \varpropto \xi^{2-\eta} \xi^{z}~~, 
\end{equation} 
where we have introduced $R \equiv \arrowvert r_i-r_j \arrowvert$. Here $z$ is the dynamical critical 
exponent. For our system we have taken $T=\Tc$, that is the critical isotherm, at constant $s=\Sc$. 
Therefore the length correlation diverges according to the magnetic field. By assuming the usual 
Ising behaviour \cite{NewmanBook}, we have
\begin{equation}\label{eq:correlationLength}
\xi \varpropto \arrowvert h \arrowvert^{-\Theta}~~~{\mathrm{, with}}~~~{\Theta=\nu{(\beta+\gamma)}^{-1}}.
\end{equation} 
From the Equations~\eref{eq:chisT} and~\eref{eq:correlationLength} we recover the static result 
$\chi^{(s)}_{_\texttt{T}} \varpropto h^{-(1-1/ \delta)}$, whilst from the Equations~\eref{eq:chiM}
and ~\eref{eq:correlationLength}, we find the dynamical critical behaviour given by
\begin{equation}\label{eq:chiMexponent}
\chi_{_\texttt{m}} \varpropto  h^{-\Delta}~~~{\mathrm{, with}}~~~\Delta=(\gamma+\nu z)(\beta \delta)^{-1}.
\end{equation} 

\begin{figure}
\includegraphics[width= .75\columnwidth]{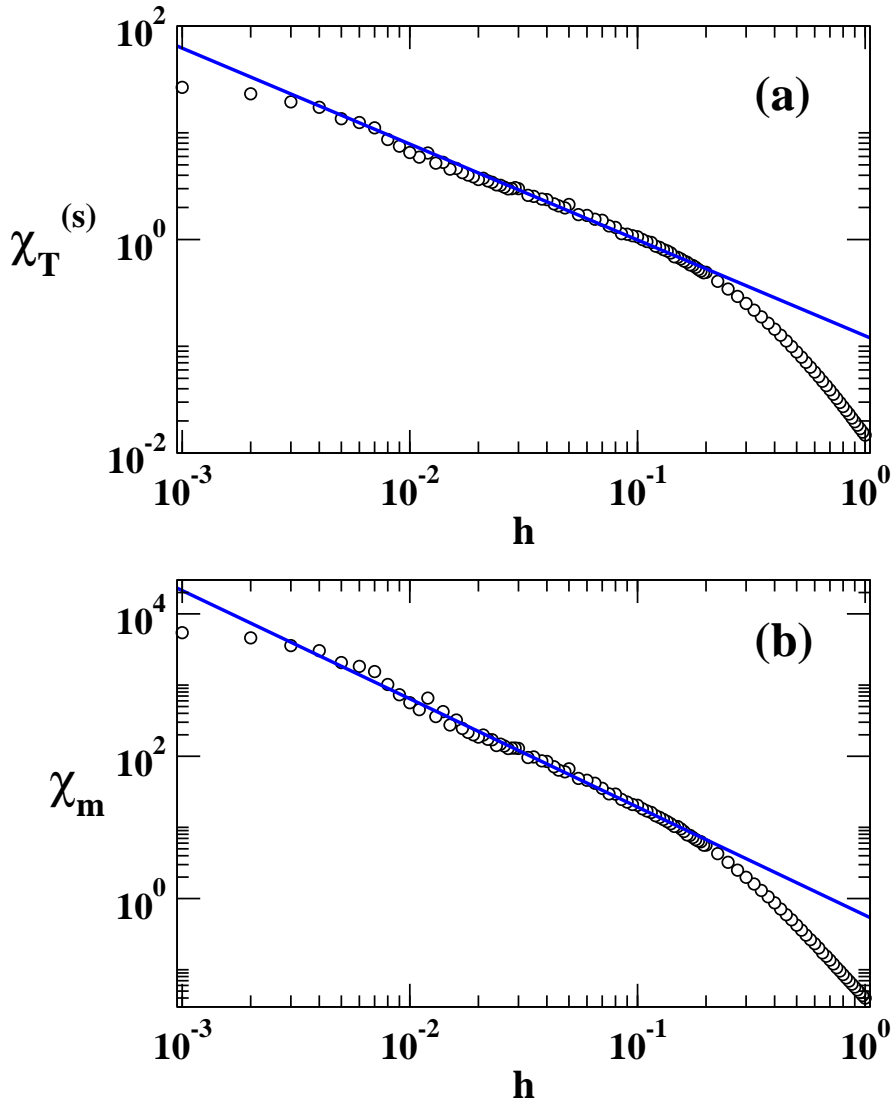}
\caption 
	{(a) Susceptibility of the instantaneous magnetization corresponding to spatial fluctuations. 
         The continuous line is a fit of a power law that gives exponent $0.90\pm 0.05$, in good agreement 
         with the exponent of the magnetic susceptibility of the 2$d$ Ising model $1-1/\delta\approx 0.933$   
         (b) Susceptibility of the integrated magnetization corresponding to spatio-temporal fluctuations. 
         The continuous line is a fit of a power law that gives exponent $1.50\pm0.05$;
        }
\label{Fig:susceptibilities}
\end{figure}

\Fref{Fig:susceptibilities} shows the results of the Monte Carlo simulation of the Ising 
chain with a magnetic field for $5\times10^6$ generated trajectories ($N=64$ and $\tobs=300$). 
A clear behaviour emerges for external fields of moderate intensity, that is $0.004 \lesssim h$.
We have found that the results for different times $\tobs=160,200,300$ (and different number of 
spins $N=64,100$) are the same within error bars. For these values of the magnetic field there 
are no finite size effects. The stationary thermodynamic behaviour is observed.
In other words, for weak field  $h \lesssim 0.004$, there are finite size effects observed, caused by 
limited $\tobs$ and lattice size $N$. Any attempt to study the phase space region with a small external 
field ($h \lesssim 0.004$) would require extensive simulation efforts which are beyond the scope of 
this paper.

\Fref{Fig:susceptibilities}(a) shows the {\it static} susceptibility $\chi^{(s)}_{_\texttt{T}}$.
We can see that a power law fits in the region of the magnetic field, $0.004 \lesssim h \lesssim 0.2$,
with a fitted exponent $0.90 \pm 0.05$. For higher field ($ 0.2 \lesssim h$) there is a deviation of 
the power law regime, as expected far from the critical point. This value is consistent with the value 
of the 2$d$ Ising model given by the~\Eref{eq:Delta}. 
On the other hand,~\fref{Fig:susceptibilities}(b) shows the susceptibility $\chi_{_\texttt{m}}$.
The behaviour of the static susceptibility is qualitatively analogous to  $\chi_{_\texttt{m}}$. 
However the power law fit within the same region of the magnetic field, $0.004 \lesssim h \lesssim 0.2$, 
gives a different exponent, that is $\Delta = 1.50 \pm 0.05$. Using the relationship given 
by~\Eref{eq:chiMexponent} and assuming the Ising universality, we can conclude that our 
simulation gives us an estimation of the dynamical critical exponent, which is $z=1.06 \pm 0.09$. 
This value is compatible with $z=1$. 

This results shows that through the application of the magnetic field to the Ising chain using 
the $s$-ensemble, the spatial correlations involved are similar to those present in the 2$d$ Ising 
model with field near the critical point, while the response of the integrated magnetization 
is governed by a (greater) exponent dependent on the dynamical exponent $z$.  

In the system presented here, the static criticality of the 2$d$ Ising model universality 
class appears in different ways. For the energy and activity criticality arises in the space-time 
plane (see figures \ref{Fig:energy} and~\ref{Fig:chi_k}) and is uncovered by means of finite size 
effects, whereas for simulations with external magnetic field it appears only in the spatial axis 
(see~\fref{Fig:susceptibilities}(a)) and in the thermodynamic limit.

\subsection{Dynamic first order transition}\label{magnetic_properties}

We argued in~\sref{section:phase-diagram}, that in the presence of an external field, 
the spontaneous broken symmetry should give rise to a first order phase transition in the ferromagnetic
phase. So that, for isothermal conditions, a first order phase transition is expected for 
the biasing field $s$ greater than the critical value $\Sc(T_0)$ (see~\fref{Fig:phase-diagram}(c)). 
Illustrating the dynamic behaviour, including an external magnetic field analytically is non-trivial. 
This is due to the fact that the mapping of the master equation onto the quantum problem will give 
rise to highly non-linear terms. In the following we will address this problem numerically. 
\begin{figure}
\includegraphics[width = .9\columnwidth]{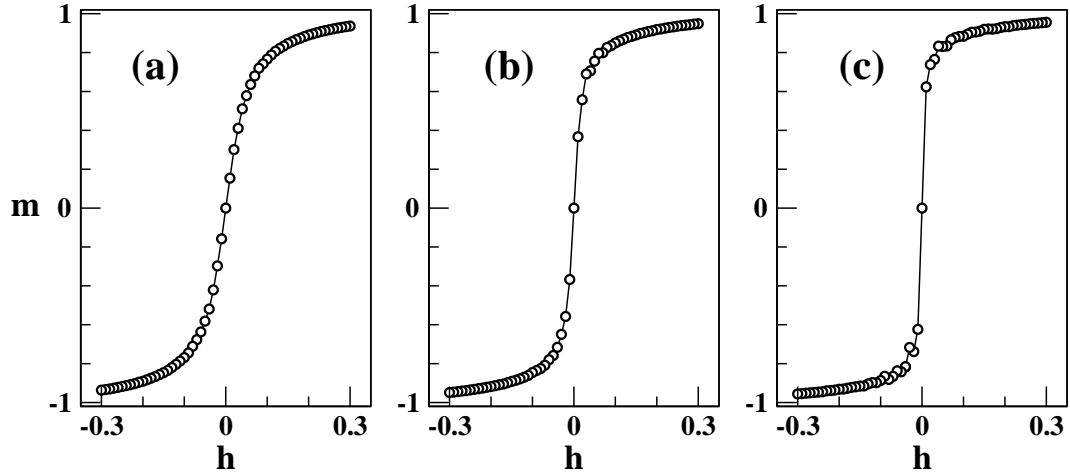}
\caption 
	{Stationary results for the magnetization of the Ising chain with magnetic field
         using the $s$-ensemble for different values of the constant biasing field $s$.
         The temperature $T$ is constant ($T=1$).
         (a) magnetization for the subcritical behaviour, that is in the paramagnetic phase $s=0.02<\Sc$,
         (b) results for the critical values of $s=\Sc=0.036~635~37$, and 
         (c) results obtained for the supercritical behaviour, that is in the ferromagnetic phase $s=0.05>\Sc$.
	}
\label{Fig:campo}
\end{figure}

We consider the parameter space $\{T,s,h\}$ with focus on an isothermal plane ($T=1$). 
In the  $h=0$ plane, two phases are present; a paramagnetic phase for $s<\Sc \approx 0.037$, and 
for $s>\Sc \approx 0.037$ a ferromagnetic phase, as depicted in the schematic 
of~\fref{Fig:phase-diagram}. \Fref{Fig:campo} shows the results of the Monte Carlos simulations
including an external magnetic field, that is the magnetization in terms of $h$ for different values 
of the biasing field $s$. In this case we have used a system of $N=200$ spins, and we generated
$10^6$ trajectories. In~\fref{Fig:campo}(a) a smooth curve $m(h)$, in the vicinity of $h=0$,
for a low biasing field $s=0.02$ can be seen, this is the paramagnetic phase. Whereas~\fref{Fig:campo}(c) 
shows a sharp transition, at $h=0$, for a supercritical value of biasing the field $s=0.05$, this is the 
ferromagnetic phase. For completeness~\fref{Fig:campo}(b) illustrates the behaviour at the critical 
point $\Sc=0.036~635~37$. The behaviour of $s$ is equivalent to that of the inverse temperature 
$\beta$ for the thermodynamic fluctuations of the Ising ferromagnet.  
Therefore, we expect that the field-driven first order phase transition ends at the critical point 
$s=\Sc$ ($T=1, h=0$) which is the limit of the ferromagnetic phase ($s>\Sc$). 

\begin{figure}
\includegraphics[width = .9 \columnwidth]{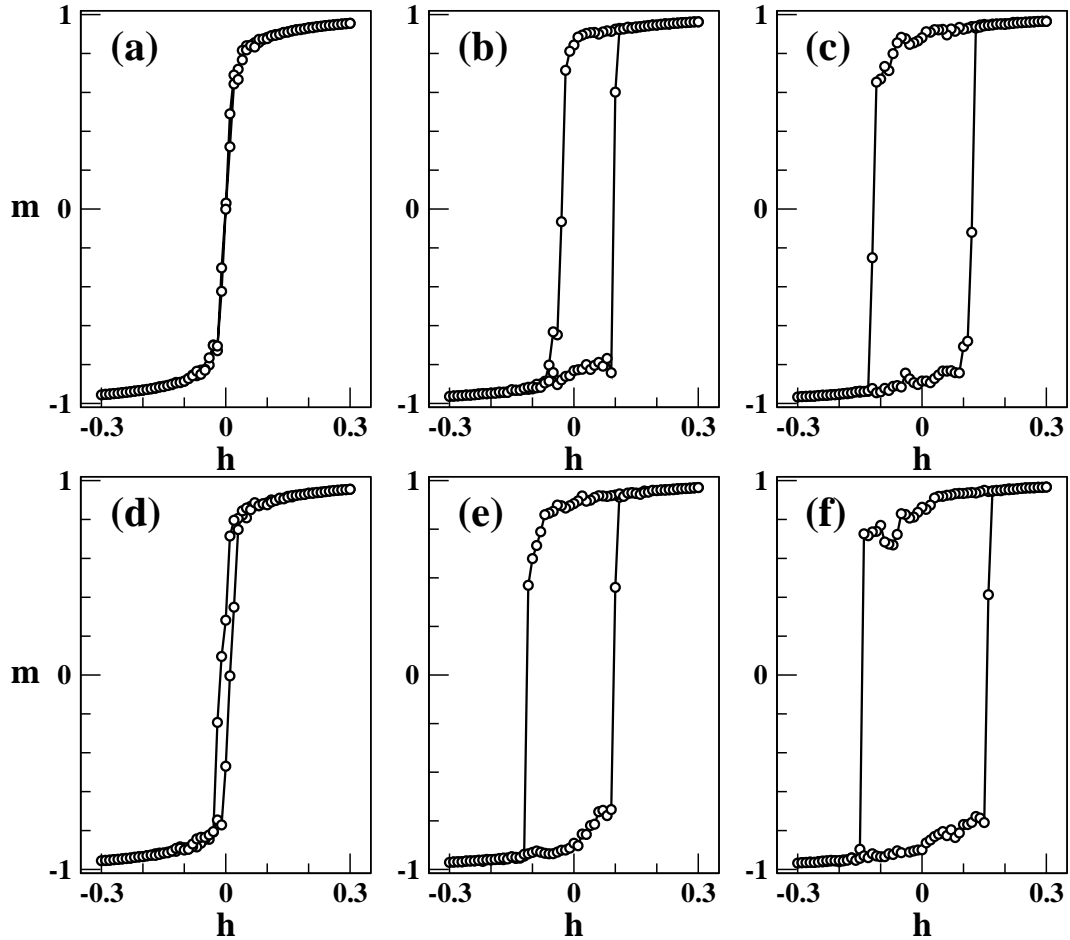}
\caption 
	{Magnetization versus the magnetic field in the form of hysteresis loops at constant temperature $T=1$
         around the first order transition at $h=0$. Also the biasing field $s$ is fixed (all of them 
         supercritical values): 
         for (a) and (d) $s=0.05$, 
         for (b) and (e) $s=0.07$, and 
         for (c) and (f) $s=0.08$.   
         In the top (bottom) panel we have used an observational time $\tobs=60$ ($\tobs=80$).
         The area of the loops increases with $\tobs$, and with $s$. More details in the text.
         }
\label{Fig:fig7}
\end{figure}

In order to capture this behaviour further, hysteretic loops were studied. The simulations 
were started from a representative trajectory of the stationary distribution, then the external 
field is perturbed by $h  \pm \Delta h$  (with $\Delta h \sim 0.01$). 
In order to reach a stationary state for the new value of $h$, $n_{\mathrm{relax}}$ steps in the 
TPS algorithm are taken for equilibration. Then measurements of the observables are taken for
$n_{\mathrm{obs}}$ steps.~\Fref{Fig:fig7} shows the resulting loops of the magnetization 
versus the magnetic field. We used a system size $N=200$ and $n_{\mathrm{relax}}=10^2$ 
TPS moves were attempted, in order to achieve relaxation and the hysteresis was studied using 
$n_{\mathrm{obs}}=10^5$ realizations.

The top panel of~\fref{Fig:fig7} depicts the hysteresis loop with increasing supercritical
$s$ with a trajectory length of $\tobs=60$, that is $s=0.05$ for (a), $s=0.07$ for (b), and $s=0.08$ 
for (c). The bottom panel of~\fref{Fig:fig7}, again shows the hysteresis loops, but in this 
case the observational time of the trajectory was increased to $\tobs=80$. The values for $s$ are 
retained ($s=0.05$ for (d), $s=0.07$ for (e), and $s=0.08$ for (f)). 
Clearly, the area of the loops increases with $\tobs$ as well as with $s$. This is 
because as $s$ increases the metastability of the system increases, therefore the loop area 
also increases. The behaviour with respect to $\tobs$ can be treated in terms of an increase in volume, 
which means the relaxation is slower and therefore an increased loop size is observed.  
This behaviour supports that the dashed line in the~\fref{Fig:phase-diagram}(c), 
and the surface below of the continuous line in the~\fref{Fig:phase-diagram}(a), 
are made of first order phase transition points.

\section{Kawasaki dynamics}\label{section:Kw}

So far we have looked at dynamical models with non-conserved magnetization \cite{Bodineau2008,Jack2010}, 
which in the case of the classic thermodynamics, would serve as an order parameter. 
These are A type models or non-conserved order parameter \cite{Bray2002}. In the particular case of the Glauber 
dynamics it has been shown, and we have verified that in the dynamic exploration of this model, 
a line of critical points in the phase space can be found. We expect this kind of behaviour 
for model A type dynamics. 

On other hand, it is interesting to ask about the role of the the underlying dynamics on the observed 
\textit{dynamical} phase diagram. In order to illuminate this point, let us now consider a model 
belonging to the second class of dynamics, where the magnetization is conserved (conserved order parameter). 
We have realized Monte Carlo simulations using Kawaksaki dynamics. This dynamics conserves 
the overall magnetization and consists of exchanging neighbouring spins, 
rather than flipping them. The acceptance function for undergoing a spin exchange is chosen 
to be the Metropolis criterion 

\begin{equation}
P_{\mathrm{accept}} = \mathrm{Min}(1, e^{-\Delta E/k_BT})~.
\end{equation}
where $\Delta E$ is the increment in the energy due to the exchange.
\begin{figure}
\includegraphics[width= .65\columnwidth]{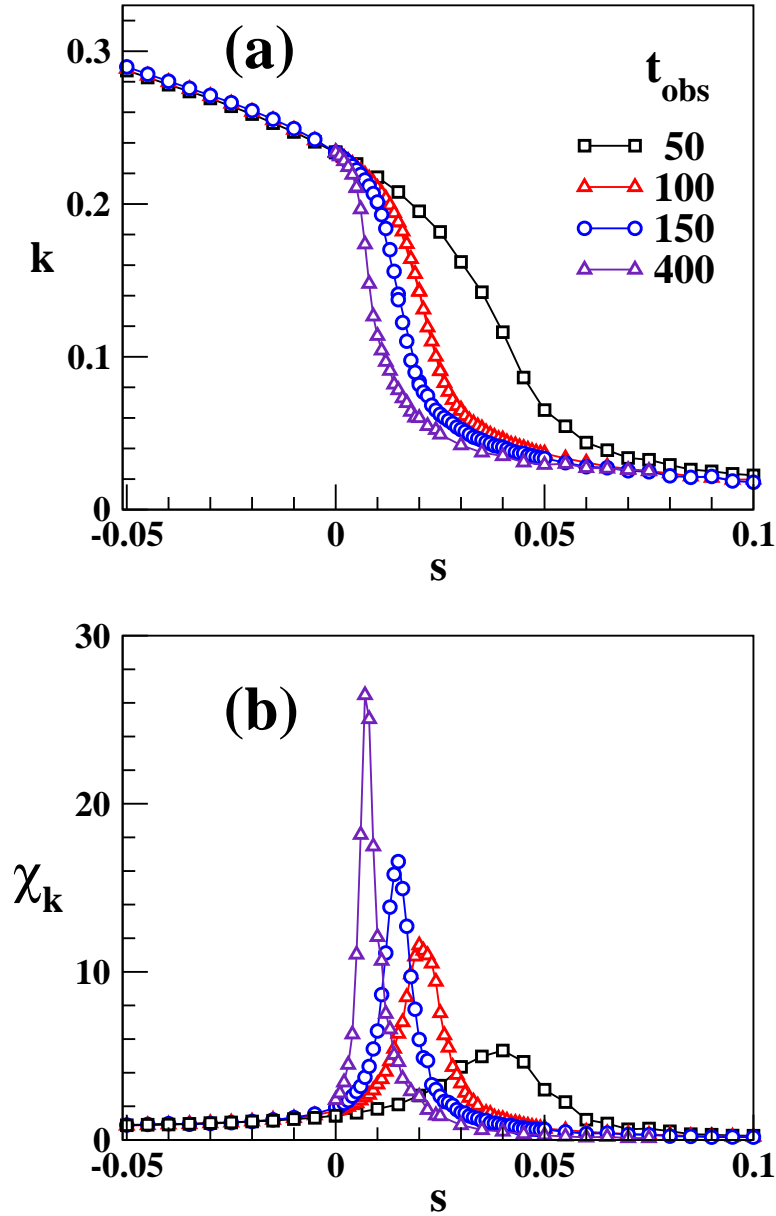}
\caption 
	{Results of the Monte Carlo simulations using the $s$-ensemble with Kawasaki dynamics
         taking the temperature constant ($T=3$). When the observational time $\tobs$ is varied,
         as indicated in (a), finite size effects are observed.
         (a) Activity versus the biasing field $s$. (b) susceptibility of the activity as a function
	 of $s$. 
         }
\label{Fig:kwactivity}
\end{figure}

\begin{figure}
\includegraphics[width= .65\columnwidth]{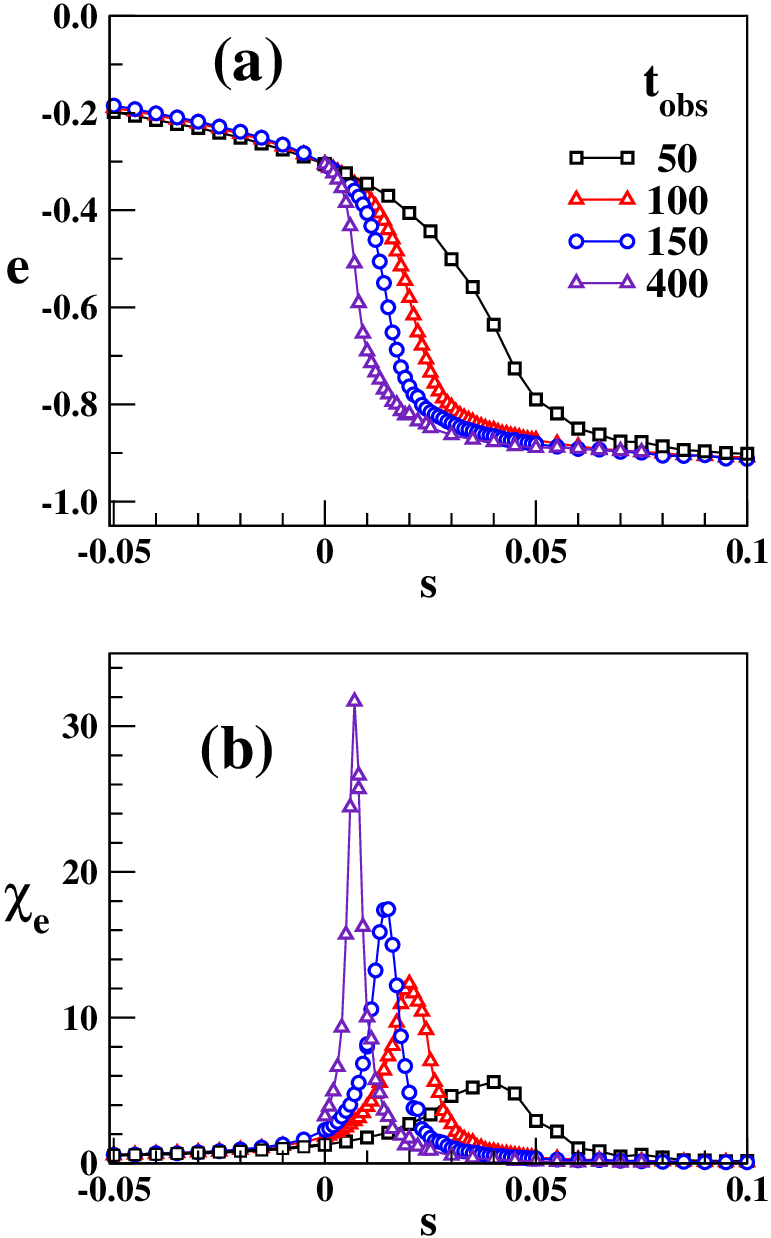}
\caption 
	{Results of the Monte Carlo simulations using the $s$-ensemble with Kawasaki dynamics
         taking the temperature constant ($T=3$). When the observational time $\tobs$ is varied,
         as indicated in (a), finite size effects are observed.
         (a) Energy versus the biasing field $s$. (b) susceptibility of the energy as a function
	 of $s$. 
         }
\label{Fig:kwenergy}
\end{figure}

In~\fref{Fig:kwactivity} we show the results of the Monte Carlo simulations using the $s$-ensemble 
with Kawaksaki dynamics for the Ising chain. These simulation were realized at constant 
temperature $T=3$ for a spin chain of size $N=64$ and zero magnetization, i.e. equal populations of spins up 
and down.~\Fref{Fig:kwactivity}(a) and~\fref{Fig:kwactivity}(b) show the activity and the susceptibility 
of the activity, given by~\Eref{Eq:activitysusc}, respectively versus the biasing field $s$ for different 
$\tobs$. For low values of $s$ we can observe an active phase which becomes inactive for higher values of $s$. 
Also, when the observational time is increased, the jump becomes more abrupt and consequently its 
susceptibility shows an increasingly sharp peak. Again this is the sign of a phase transition and 
suggest a good finite size scaling behaviour through increasing the length scale, i.e. $\tobs$,
in the same way was done in~\sref{section:Glauber}.  

\Fref{Fig:kwenergy} shows the behaviour of the integrated energy with respect to $s$ and its 
susceptibility defined by~\Eref{Eq:energysusc}. This behaviour is qualitatively the same as 
the behaviour of the activity. We can understand this similarity qualitatively.
The energy is proportional to number of domain walls (that is the number of the pair of first 
neighbours with different spins). Using the Kawasaki dynamics we require two neighbours sites
with different spins, in order to make a spin flip, thus the same factor governs both 
the energy and activity. 

The figures~\ref{Fig:kwactivity} and~\ref{Fig:kwenergy} suggest, that the observed transitions 
are first order transitions. Generally finite size scaling of a first order transition~\cite{Landau2005}
predicts scaling in terms of the volume $V=L^D$. According to this theory, for a finite volume 
$V$, we can define an effective critical point $\Sc$ as value of $s$ at which the susceptibilities 
have the maximum $\chi^{max}$. In this way, the following scaling behaviour is expected
\begin{equation}\label{eq:FSSFOextrapolation}
\Sc(\tobs) = \Sc^{(\infty)} + A V^{-1} ~~ ,
\end{equation}
where $\Sc^{(\infty)}$ is the thermodynamic limit $\Sc(\tobs\rightarrow\infty)$, and
\begin{equation}\label{eq:FSSFOchimax}
\chi^{max} \varpropto V ~~ .
\end{equation}
In our case, the volume in the space-time is given by $V=N\tobs$. By fixing the number of spins $N$,
we can write out $V = \tobs$ in Equations \eref{eq:FSSFOextrapolation} and \eref{eq:FSSFOchimax}.

\begin{figure}
\includegraphics[width= .7\columnwidth]{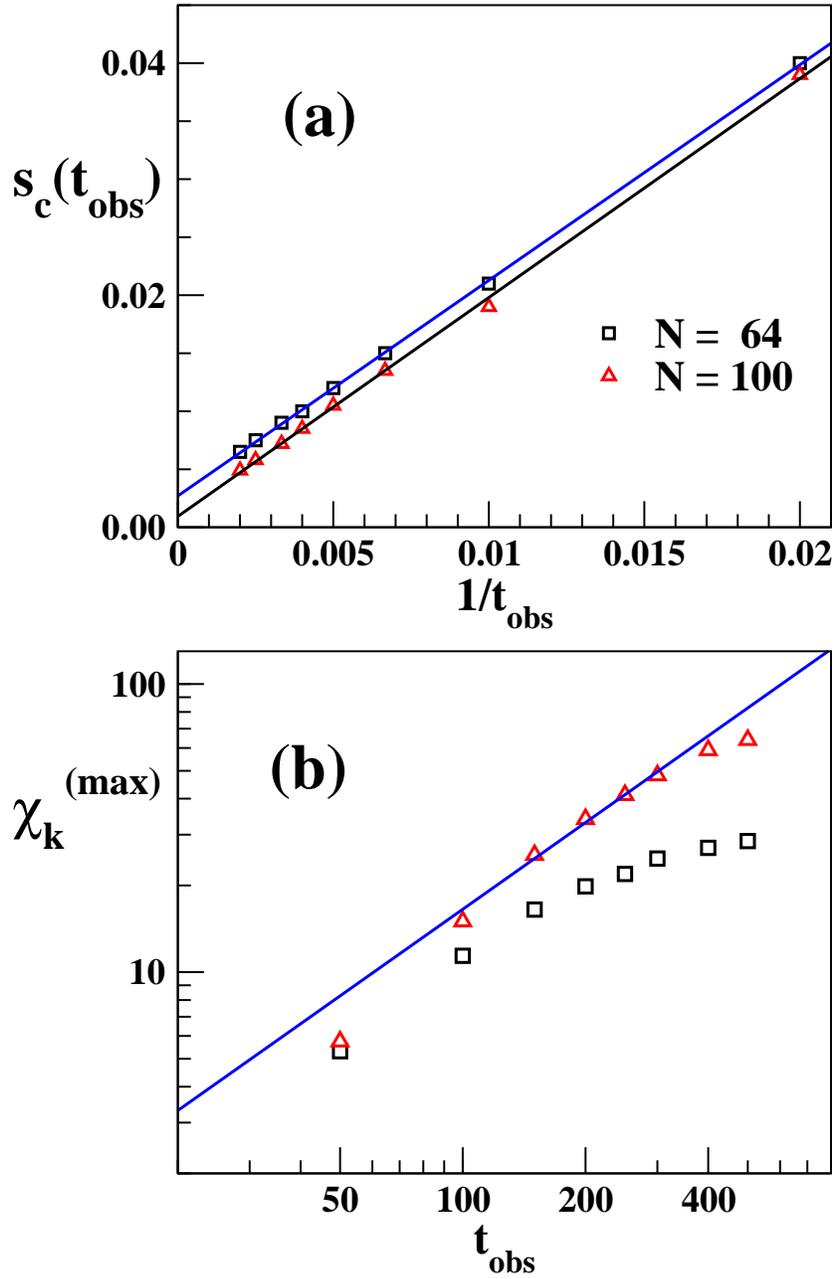}
\caption 
	{Results of the Monte Carlo simulations using the $s$-ensemble with Kawasaki dynamics
         taking the temperature constant ($T=3$). Black Squares are results obtained from the 
         data of~\fref{Fig:kwactivity} for $N=64$. For comparison, red triangles are 
         data obtained from simulations with $N=100$ analogous to~\fref{Fig:kwactivity} 
         (data not showed for the sake of clarity).  
         (a) The effective critical biasing field $\Sc$, defined as the location of the peak in the
         susceptibility $\chi_{\texttt{k}}$, as a function of $1/\tobs$. The straight lines are fits used 
         for the extrapolations of $\Sc$ to the limit $\tobs\rightarrow \infty$, by means of the scaling 
         valid for first order transitions given by~\Eref{eq:FSSFOextrapolation}.
         The result is $\Sc=0.003\pm0.001$ ($\Sc=0.001\pm0.001$) for $N=64$ ($N=100$). 
         (b) height of the peak of the susceptibilities of the activity for the same cases that in (a).
         The straight line is the expected power law $\varpropto\tobs$ from the finite size 
         theory for first order transitions. More details in the text. 
        }
\label{Fig:kwpeaks}
\end{figure}

\begin{figure}
\includegraphics[width= .9\columnwidth]{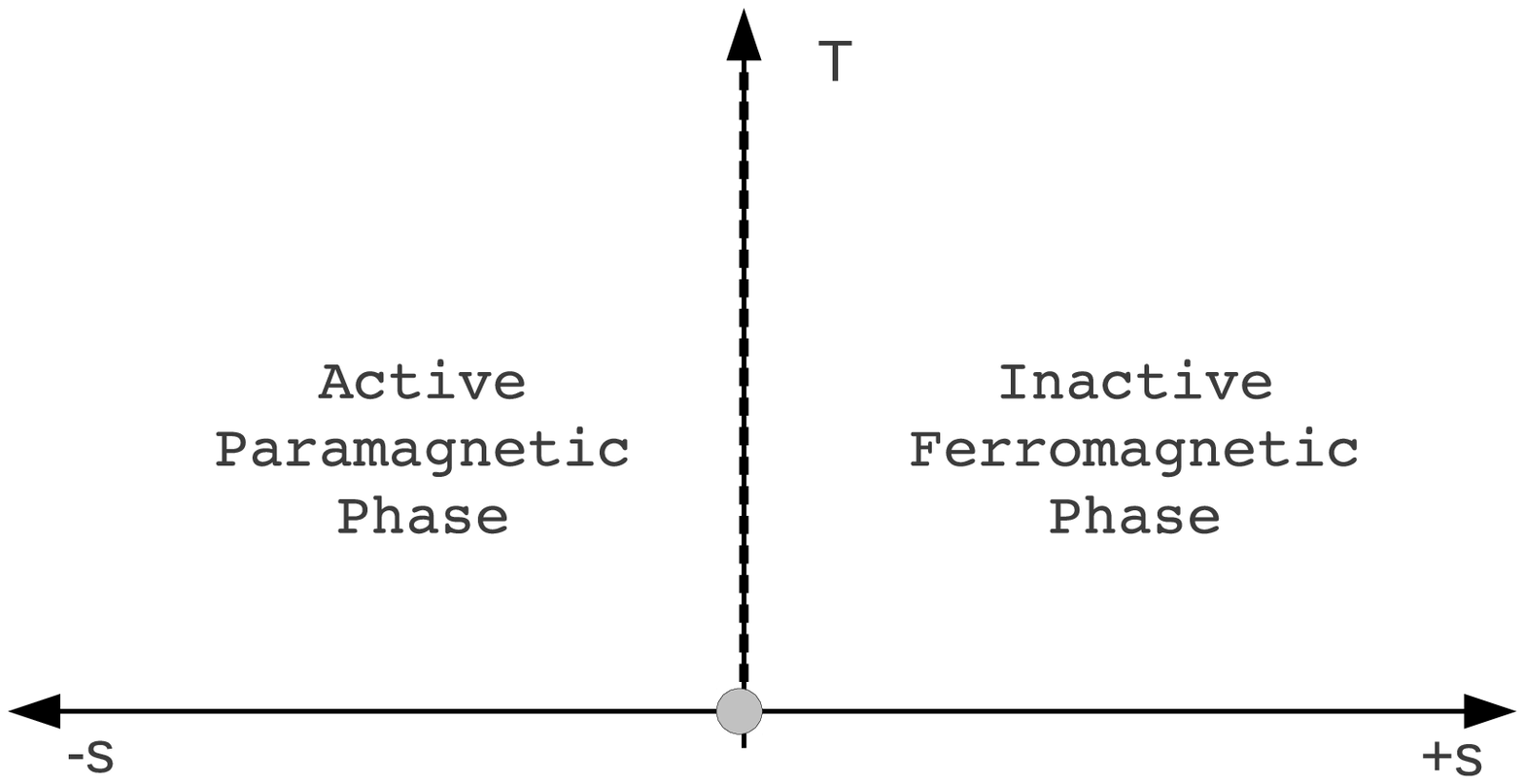}
\caption 
	{Phase diagram of the Ising chain with Kawaksaki dynamics. The dashed line for $s=0$
         represents points of a first order phase transition between the indicated phases.
        }
\label{Fig:kwphasediagram}
\end{figure}

\Fref{Fig:kwpeaks} illustrates that the finite size analysis holds for the activity 
(the same behaviour is also observed for the energy).
In~\fref{Fig:kwpeaks}(a) the location $\Sc$ of the maximum $\chi_{\texttt{k}}$ is plotted for two Ising 
chains of length  $N=64$ and $N=100$ against $\tobs^{-1}$.
The effective critical value $\Sc$ decreases with increasing $\tobs$ and with increasing $N$.
Using~\Eref{eq:FSSFOextrapolation} to calculate the extrapolated value to $\tobs\rightarrow \infty$
we obtain for $N=64$ the value $\Sc=0.003\pm0.001$, and $\Sc=0.001\pm0.001$, for $N=100$ 
(the same results are obtained from the energy analysis).

\Fref{Fig:kwpeaks}(b) shows the height of the peaks drawn from the susceptibility of the activity 
versus the observational time. The continuous line in this figure is a power law given by 
the~\Eref{eq:FSSFOchimax}. For the smaller system size the power law
is not observed at all. For the larger size we can see that there is a narrow region, for intermediate 
values of $\tobs$, where the power law holds. These result shows that for small $\tobs$ there are 
boundary effects in the time of the trajectories, and for longer $\tobs$ there are boundary effects 
in the space of the spin chain. These rounding effects, make it difficult to observe the scaling behaviour 
given by~\Eref{eq:FSSFOchimax} for the sizes and observational times used in this paper.

The previous analysis of figures~\ref{Fig:kwpeaks}(a) and~\ref{Fig:kwpeaks}(b) indicates a 
first order phase transition, between an active phase (for $s$ negative) and an inactive phase
(for $s$ positive) and a transition point at $\Sc = 0$.  This space-time phase structure, of 
coexistence of active and inactive dynamical phases at $s=0$, is similar to that of idealised
kinetically constrained models \cite{Garrahan2007,Garrahan2009}.  
 
It is worth mentioning that the activity and the energy, in the inactive phase 
(see figures~\ref{Fig:kwactivity}(a) and~\ref{Fig:kwenergy}(a)) are $\sim1/N$ due to 
finite size effects. In fact the dynamics implies that there always are, at least two domain walls,
which separate the up and down pointing spin domains in order to conserve zero magnetization. 
Because of this constraint the minimum of the energy per site is given by $e-e_0=2/N$, where $e_0$ is 
the unconstrained energy corresponding to the ground state. For increasing $N$ and increasing $\tobs$ 
the energy in this phase tends to this characteristic value, and consequently the activity is of the same order.
This behaviour shows that the inactive phase is a ferromagnet, where the lattice phase separates into two domains,
one consisting of up spins and the other of down spins. Such state will have zero activity in the 
thermodynamic limit. 

We have performed the same analysis for other temperatures, and we conclude that this first order dynamical
transition between a paramagnetic/active phase and a ferromagnetic/inactive phase at $\Sc=0$ is present 
for all temperatures considered ($T=1,2,3$). At low temperatures, the accessibility of the transition becomes 
more difficult because the value of activity will be much smaller at lower temperatures and it will be zero in 
the inactive phase, so that the difference between the two phases is too small to resolve. In the limit 
$T\rightarrow 0$ the discontinuity disappears and the transition ends in the point $s=0,T=0$. Once again, 
this is analogous to what occurs in kinetically constrained models \cite{Garrahan2007,Garrahan2009}.

\Fref{Fig:kwphasediagram} presents the phase diagram that summarizes the result discussed above. 
This phase diagram for the Ising chain using Kawasaki dynamics should be compared 
to~\fref{Fig:phase-diagram} illustrating the Glauber dynamics. These diagrams suggest 
a very different phase behaviour for Models A and B. Models B only exhibit a dynamic 
phase transition at $s=0$ between two dynamic asymmetric phases, one is an active 
paramagnet and the other is an inactive ferromagnet. This first order transition is driven by the 
biasing field $s$. For Models A, there is a surface of first order lines, between two symmetric 
ferromagnetic phases and it is driven by the magnetic field. The surface ends in a critical line, 
in the 2$d$ Ising universality at $s\neq0$. 

\section{Conclusions}\label{section:Conclusions}

In this paper we have considered how dynamical phase behaviour relates to thermodynamics in the 
context of the one-dimensional Ising model. We have studied $s$-ensembles of trajectories 
\cite{Lecomte2007,Garrahan2007,Garrahan2009,Hedges2009} by means of path sampling 
simulation methods for both Glauber and Kawasaki dynamics.  We have both confirmed the analytic
predictions of reference~\cite{Jack2010} for Glauber dynamics, and extended the space-time phase 
diagram to the case of non-zero magnetic field.  Our results show an intimate relation between 
thermodynamic and dynamic phases, but this relation depends on the details of the dynamical rules.
For both Glauber and Kawasaki dynamics we show that the paramagnet (which is the stable thermodynamic 
phase for all $T>0$ in one dimension) has a counterpart dynamical phase of high activity.  While 
thermodynamically unstable, the  static ferromagnetic phase corresponds dynamically to a phase of
low activity, and by biasing ensembles of trajectories via the counting field $s$ it is possible
to go through a dynamic/non-equilibrium phase transitions between the active paramagnetic, and the 
low-activity ferromagnetic phases. In the case of Glauber dynamics, as predicted in \cite{Jack2010},
this transition is second-order and in the $2d$ Ising model universality class, a fact that we 
verified numerically by detailed finite size (and time) scaling. Furthermore, with the inclusion 
of a magnetic field $h$ we were able to show that the area below the line of critical points in the
$T-s$ plane at $h=0$ is actually  a surface of first order transitions between low-activity ferromagnetic 
phases of positive or negative overall magnetization. This is a very interesting example of dimensional
reduction: ensembles of trajectories of the 1$d$ Glauber Ising model have statistical properties similar
to ensembles of configurations of the equilibrium 2$d$ Ising model.

We also considered the 1$d$ Ising model under Kawasaki dynamics. In this case the transition line between 
the active/paramagnet and the inactive/ferromagnet lies on the $s=0$ axis, a situation analogous to 
that of kinetically constrained models of glasses \cite{Garrahan2007,Garrahan2009}. 
This can be understood by the fact that under Kawasaki dynamics spin-exchange leads to 
transitions in terms of domain walls where there is at least one domain wall present either 
before or after the transition: the presence of this persistent domain wall plays the role of a 
kinetic constraint.

The aim of this paper was to shed light on the connection between structural or thermodynamic 
phase structure, and space-time or dynamic phase structure as revealed by the large-deviation 
method of trajectories.  The natural next, and more challenging, step in this programme will be 
to perform a similar analysis on prototypical systems which display thermodynamic transitions at
non-zero temperature, such as the 2$d$ Ising model.

\ack
We wish to thank Robert Jack and Fr\'ed\'eric van Wijland for helpful suggestions 
and comments. The authors are also grateful to Argentinian Science Agency CONICET, 
and British Council France Alliance Project 09.013, for their financial support.

\section*{References}

\end{document}